# Electric properties of carbon nano-onion/polyaniline composites: a combined electric modulus and ac conductivity study

Anthony N. Papathanassiou [a,*], Olena Mykhailiv[b], Luis Echegoyen [c], Ilias Sakellis[a] and Marta E. Plonska-Brzezinska [b,*]

(a) National and Kaposdistrian University of Athens, Physics Department, Solid State Physics Section, Panepistimiopolis, GR 15784 Athens, Greece
(b) University of Bialystok, Institute of Chemistry, Hurtowa 1, 15-399 Bialystok, Poland
(c) University of Texas at El Paso, Department of Chemistry, 500 W. University Ave., El Paso, TX 79968 USA

**Abstract**

The complex electric modulus and the ac conductivity of carbon nano-onion/polyaniline composites were studied from 1 mHz to 1 MHz at isothermal conditions ranging from 15 K to room temperature. The temperature dependence of the electric modulus and the dc conductivity analyses indicate a couple of hopping mechanisms. The distinction between thermally activated processes and the determination of cross-over temperature were achieved by exploring the temperature dependence of the fractional exponent of the dispersive ac conductivity and the bifurcation of the scaled ac conductivity isotherms. The results are analyzed by combining the granular metal model(inter-grain charge tunneling of extended electron states located within  mesoscopic highly conducting polyaniline grains)  and a 3D Mott variable range hopping model (phonon assisted tunneling within the carbon nano-onions and clusters).

**Corresponding authors;** e-mail address: antpapa@phys.uoa.gr, mplonska@uwb.edu.pl

# 1. Introduction

Carbon nano-onions (CNOs) are spherical multi-layer carbon structures composed of concentric fullerenes similar to Russian dolls. The effective surface areas of CNOs are considerably larger than those of other allotropes of carbons, such as carbon nanotubes, graphene, graphite, etc [1, 2]. Electrically percolating composites consisting of a polymer matrix filled with carbon nanostructures have been studied to obtain optimum electrical, electromagnetic, mechanical and thermal properties [3]. Recently, the effect of CNO loading on the dielectric and electric properties of composites was investigated [4, 5, 6]. For different composites, the concentrations of the carbon nano-particles and their percolation threshold (i.e., the critical concentration for which a continuous conduction network is available to electric charge carriers) were studied. Interestingly, it was found that the presence of CNOs induced huge capacitance effects (visualized as intense static dielectric constant values), coexisting with the percolation network [7]. The investigation of the role of CNOs as composite fillers at high weight ratio is largely unexplored and requires further experimental and theoretical work.

The CNO and conducting polymer (CP) composites are highly conducting due to the conjugated bond structures of the CP and the intrinsic electronic properties of the CNOs. Quantum mechanical tunnelling assisted by phonons occurs in the inhomogeneous structure of the CP and the defected structure of the CNO. The composites are promising efficient electrode materials for energy storage super-capacitors [8, 9] and for biocompatible electrodes [10, 11, 12]. The multi-layered carbon arrays yield large capacitance values, a property related to charge storage, while a conductivity percolation network permits charge transfer to occur along the specimen. Alternatively, the water-dispersed composites may form materials with hydrophilic hydrogels, yielding soft electrodes that are compatible with tissues or organs in order to emit or collect electric signals (charge flow) stemming from biological processes. An insight at the microscopic scale IS crucial for technological applications. The knowledge of electric charge transport mechanism at the microscopic scale, facilitates optimization of the carbon nano-onion/polyaniline (CNO/PANI) composite properties as super-capacitor electrode materials, by tuning the efficiency of individual microscopic conduction mechanisms by proper physico-chemical treatments (i.e., by doping or de-doping, the structural and electrical modification etc.).

Polyaniline exists in a variety of forms that differ in chemical and physical properties. The three principal forms are: leucoemeraldine (LE, benzenoid structure), emeraldine (EM, combination of quinoid and benzenoid structure), and pernigraniline (PE, quinoid structure) [13, 14]. The colours change during the redox transformations of polymer: yellow ↔ green ↔ blue [15], depending on the polymerization method. The green protonated EM exhibits semiconductor behaviour [16], where protonated polyaniline can be converted to a non-conducting blue emeraldine base form [17]. Recently, we have focused on the synthesis of the CNO/PANI composites including the green protonated EM form, using an *in-situ* chemical oxidative polymerization method. The procedure for the preparation of the CNO/PANI composites has already been reported in previous studies [18, 19]. Initially, CNOs were functionalized with phenyleneamine-terminated groups (4-aminobenzoic acid). This functionalization reaction led to a pronounced increase of the solubility of the CNOs in the protic solvents. The green protonated conducting emeraldine form of polyaniline was attached to the terminal groups in an aqueous acidic solution. The reaction led to the CNO/PANI composites with core-shell structures, which was confirmed using transmission electron microscopy studies [19]. The modified carbon nano-onion structures had the particles with diameters between 8-20 nm [19]. The size is indicative of the formation of a relatively thin and uniform–layer with a 1-4 nm thickness around the carbon nano-onions.

To study the chemical composition of the CNO/PANI composites, we applied Broadband Dielectric Spectroscopy (BDS) in the frequency range from 1 mHz to 1 MHz at temperatures from 15 K to room temperature. Compressed disk-shaped pellets of 4 mm diameter and 0.4 mm thickness were prepared with an IR press. Two thin indium foils were mechanically attached and squeezed on the parallel surfaces (comparative BDS scans at room temperature were done with various electrode materials such as bronze, gold, platinum, etc to test electrode effects). Indium, being a soft metal, can make good contact with the specimens' surfaces [7]. The indium-specimen-indium 'sandwich' was placed in a capacitor type sample holder of a high-vacuum liquid helium cryostat (ROK, Leybold-Hereaeus) operating from 15 K to room temperature. Temperatures were stabilized by a LTC 60 temperature controller. Complex permittivity measurements were recorded at frequencies ranging from 1 mHz to 1 MHz with a Solartron SI 1260 Gain-Phase Frequency Response Analyzer, and a Broadband Dielectric Converter (Novocontrol). The data acquisition system was monitored by the WinDeta (Novocontrol) software. [7, 20, 21]. The temperature evolution of the complex electric modulus and the ac conductivity were studied. The use of the electric modulus investigation suppresses undesirable capacitance effects due to electrode contacts and provides a clear view of dc conduction and dipole relaxation. Ac conductivity vs. frequency measurements trace different spatio-temporal scales of electric charge flow and the formation of the percolation cluster. Isotherms ranging over a broad range of temperature can unravel different conductivity mechanisms (since each one of them, exhibit different temperature dependences).

## 2. Theoretical background

The complex electric modulus M* is defined as the inverse of the complex permittivity $\varepsilon^*$ [22, 23, 24]:

$$M^*(\omega) = 1/\varepsilon^*(\omega) \qquad (1)$$

where ω is the angular frequency ($\omega = 2\pi f$; f denotes the frequency) [and $i \equiv \sqrt{-1}$]. The real and imaginary parts of M* are correlated with the corresponding ones for $\varepsilon^*$ through the following relations:

$$M'(\omega) = \frac{\varepsilon'(\omega)}{[\varepsilon'(\omega)]^2 + [\varepsilon''(\omega)]^2} \qquad (2)$$

$$M''(\omega) = \frac{\varepsilon''(\omega)}{[\varepsilon'(\omega)]^2 + [\varepsilon''(\omega)]^2} \qquad (3)$$

The electric modulus for a *material exhibiting dc conductivity* $\sigma_0$ (i.e., $\sigma(\omega) = \sigma_0$ and $\varepsilon_s = \varepsilon_\infty$, where $\varepsilon_s$ and $\varepsilon_\infty$ are the low (static) and high frequency limits of the relative dielectric permittivity, respectively) is:

$$M^*(\omega) = M_s \frac{i\omega\tau_\sigma}{1+i\omega\tau_\sigma} = M_s \frac{(\omega\tau_\sigma)^2}{1+(\omega\tau_\sigma)^2} + iM_s \frac{\omega\tau_\sigma}{1+(\omega\tau_\sigma)^2} \qquad (4)$$

where $M_s = 1/\varepsilon_s$ and $\tau_\sigma = \varepsilon_\infty/\sigma_0$ is the so-called 'conductivity relaxation time'. Eq. (4) yields:

$$M'(\omega) = M_s \frac{(\omega\tau_\sigma)^2}{1+(\omega\tau_\sigma)^2} \qquad (5)$$

$$M''(\omega) = M_s \frac{\omega\tau_\sigma}{1+(\omega\tau_\sigma)^2} \qquad (6)$$

A plot of $M''(\omega)$ against $\log\omega$ yields a Debye-like peak with a maximum at $\omega_{0,\sigma} = 2\pi f_{0,\sigma} = 1/\tau_\sigma$. The latter, combined with equation $\tau_\sigma = \varepsilon_\infty/\sigma_0$, yields:

$$f_{0,\sigma} = \sigma_0/(2\pi\varepsilon_\infty) \qquad (7)$$

While complex permittivity measurements may suffer from undesirable low frequency space charge capacitance contributions, the complex electric modulus formalism suppressed them and permits a clear study of dc conductivity vs. temperature by studying the temperature evolution of a conductivity relaxation peak.

*If the dc conductivity mechanism co-exists with a dielectric relaxation* mechanism (namely 'dipolar') characterized by a relaxation time $\tau_d$, the electric modulus function takes the form:

$$M^*(\omega) = M_s \frac{i\omega\tau_\sigma}{1+i\omega\tau_\sigma} + (M_\infty - M_s)\frac{i\omega\tau_d}{1+i\omega\tau_d} \qquad (8)$$

ote that, for the simple case of a Debye relaxation, $\tau_d$ is related to the *(true) dielectric relaxation time* $\tau_\varepsilon$, by $\tau_d = (M_s/M_\infty)\tau_\varepsilon$. Accordingly, dc conductivity and dielectric relaxation mechanisms contribute as distinct peaks in a $M''(\omega)$ representation.

The complex conductivity is interconnected to $\varepsilon^*(\omega)$ according to the relation $\sigma^*(\omega) = \sigma_0 + i\omega\varepsilon^*(\omega)$ [25]. The real part of the complex conductivity $\sigma'(\omega)$ relates to the electric charge flow of various spatio-temporal scales and roughly described by an empirical relation proposed earlier [26, 27]:

$$\sigma'(\omega) = \sigma_0 + A\omega^n \qquad (9)$$

where A is a constant and n is the so-called fractional exponent, which is a positive number usually equal or lower than unity (sometimes larger than 1). In the low frequency limit, $\sigma'(\omega) = \sigma_0$. Above a crossover frequency $\omega_c$, the conductivity becomes frequency-dispersive, as carriers are making better usage of shorter pathways provided [28, 29]. The temperature (and composition) dependence of the parameters of Eq. (9) were applied to determine electric charge transport dynamics [30, 31, 32]. n is related to the dimensionality of the system using percolation theory [33]. For ionic materials, n depends only on the dimensionality of the conduction system [34]. Ac conductivity isotherms of strongly disordered solids that share common dynamics usually superimpose on a single master curve, when $\sigma'(\omega)/\sigma_0$ is plotted against $\omega/\omega_c$, where $\omega_c$ is a frequency that signals the transition from the low frequency dc regime to the dispersive (frequency-dependent) one [35, 36]. The scaling function that best fits the overlapping data-sets is independent of temperature. A commonly used scaling function is that proposed by Almond and West [37]:

$$\sigma'(\omega)/\sigma_0 = 1 + (\omega/\omega_c)^N \qquad (10)$$

where N is a fractional exponent and $\omega_c$ is the angular frequency for which $\sigma_0 = A\omega_c^n$; i.e., $\omega_c = (\sigma_0/A)^{1/n}$.

## 3. Results and discussion

The imaginary part of the complex permittivity $\varepsilon''$ consists of an intense dc conductivity component and a Cole-Cole dielectric relaxation peak (Figure 1). The relaxation dynamics of the latter has been discussed in previous work [7]. The dc component appears as a peak in the M´´representation, overlapping with the dielectric relaxation peak. According to theory as determined in the preceding section, this peak is shifted towards higher frequency (beyond the highest available working frequency) relative to its location in the ε´´(f) representation. The temperature dependence of M´ and M´´are depicted in Figures 2 and 3, respectively.

The mass ratio of CNO to PANI in the composite was 1:1. Thus, a single dc conductivity measurement at a given temperature (such as room temperature) provides an effective conductivity value that contains many discrete (and unresolved) conduction contributions. The different conductivity mechanisms depend mainly on temperature. Therefore, different conductivity mechanisms can be distinguished by measuring over a broad temperature region (from 15 K to 294 K). In Figure 4, isotherms of σ´(ω) are depicted.

The dc conductivity is estimated from ac conductivity vs. frequency measurements as follows. At any temperature, the ac conductivity consists of a low frequency plateau (which is dominated by the dc conductivity $\sigma_0$). The measured $\sigma_0$ is suppressed as the temperature decreases, indicating that conduction is thermally activated. Theoretical expressions for the temperature dependence of the dc conductivity $\sigma_0(T)$, can easily modify the M´´(T) expressions, as in Eq. (7), under the condition that $\varepsilon_\infty$ is practically constant. Figure 5 indicates a linear relation between $\log\sigma_0$ and $\log f_{0,\sigma}$ (data points correspond to these temperatures of which the isotherm BDS data were collected).

*3.1 Evidence for two hopping conductivity mechanisms*

Arrhenius diagrams of a dynamic quantity (such as conductivity, relaxation time, diffusivity, etc) vs. reciprocal temperature have are often used to discriminate between different mechanisms [38] and have an estimate of the apparent activation energy value.   In Figure 6, $\sigma_0(T)$ plotted as a function of 1/kT, respectively. Fits of the data obtained using a linear combination of a temperature-independent component and two exponential constituents:

$$\sigma_0(T) = S_0 + S_1 \exp(-E_{\sigma 1}/kT) + S_2 \exp(-E_{\sigma 2}/kT) \qquad (11)$$

where $S_0$ and $F_{0,\sigma}$ denote the temperature independent terms due to coherent tunneling, and $S_1$ and $S_2$ are pre-exponential constants and $E_{\sigma 1}$ and $E_{\sigma 2}$ denote apparent activation energies, respectively. Note that it was essential to take into account the first constant term in order to fit the whole set of data points efficiently. The fitting parameters to Eq. (11) are: $S_0 = (1.1\pm0.7)\times10^{-5}$ S/cm, $S_1 = (2.3\pm0.3)\times10^{-3}$ S/cm, $S_2=(4.5\pm0.7)\times10^{-5}$ S/cm, $E_{\sigma 1} = (55\pm3)$ meV and $E_{\sigma 2} = (4\pm2)meV$. The low activation energy is attributed to hopping transport along the percolation network formed by the conductive CNOs, while the high activation energy mechanism is ascribed to hopping transport through the less conductive PANI phase. The transition temperature was around 115K. This temperature cross-over has also been observed in polymer matrices accommodating various carbon nano-structures. The complex conductivity studies in the carbon nanotubes/carbon nano-onions/polyurethane (CNT/CNO/PU) and the CNO/PU composites [5] and multiwall carbon nanotubes (MWCNT)/poly(methyl methacrylate (PMMA) [39] indicated that the dielectric permittivity and electrical conductivity are most pronouncedly reduced below 100 K.

The cross-over from one mode of conduction to another,(at 115 K) is supported by the analyses of the conductivity curves. Eq. (7) was fitted to the σ´(f) data points plotted in Figure 4, and the fractional exponent n was obtained. For percolative transport along a percolation network in a disordered material, the fractional exponent increases from its room temperature value to higher values systematically when temperature is reduced (or remains constant) [40]. In Figure 7, where n(T) is depicted, we observe that the high temperature n(T) values are around 0.96, while the low temperature ones around 0.81. This indicated that two different modes of electric charge transport operate, in agreement with the findings of the electric modulus and the dc conductivity dependencies. The transition temperature was around 115 K. In Figure 8, the ac conductivity isotherms were scaled following the Almond-West approach. Scaled plots below and above 115K superimpose on two different master curves, respectively, appearing as a 'bifurcation' of the right hand side of Figure9. The transition temperature signs a 'bifurcation' in the scaling representation; each branch evidences about a change in conduction dynamics.

*3.2 Microscopic models*

Arrhenius plots and the temperature dependence of the fractional exponent of the frequency-dependent ac conductivity indicate a cross-over temperature for electric charge transport at 115K. Therefore, at least two components contribute to fit the temperature dependence of the dc conductivity and the conductivity relaxation frequency.

The structure of conducting PANI is inhomogeneous resembling that of a granular metal. Coupling between PANI chains form highly conducting regions or bundles, in which, wave functions are extended in three dimensions [41]. Charge tunneling occurs between mesoscopic metallic islands hosted by a less conductive polyaniline environment, in which the charging energy is high [42, 43]. For low electric fields, dc conduction proceeds by phonon assisted tunneling through the effective potential barrier set by the insulating host matrix. The temperature dependence of the dc conductivity is described by the granular metal (GM) model [44, 45]:

$$\sigma_{0,GM}(T) = A_{GM} \exp(-(T_{0,GM}/T)^{1/2}) \qquad (12)$$

where $A_{GM}$ is a constant and $T_{0,GM}$ is a parameter related to the height of the effective potential barrier. The nominator $T_{0,GM}$ is determined by the size of conducing grains (or clusters) and separation among them according to [46]:

$$T_{0,GM} = \frac{8U}{k} \frac{(s/d)^2}{0.5+(s/d)} \qquad (13)$$

where d is the average diameter of the conducting grains, s the average length of the depleted zone between the grains and U the dynamic energy of the Coulomb repulsion of two electrons at a distance equal to the size of the monomer, which is roughly equal to 2 eV [45, 46].

Eqs. (7) and (12) yield:

$$f_{0,\sigma,GM}(T) = A_{0,\sigma,GM} \exp(-(T_{0,GM}/T)^{1/2}) \qquad (14)$$

where $A_{0,\sigma,GM} \equiv A_{GM}/(2\pi\varepsilon_\infty)$.

The Mott-Davis variable range hopping (VRH) model is commonly used for describing the temperature dependence of the electrical conductivity in carbon allotropes [46], whereas a $T^{-\gamma}$ dependence is predicted ($\gamma$ = 1/2, 1/3 and ¼ for 1D, 2D and 3D electric charge transport, respectively). The dimensionality is related to theshape and symmetry of the carbon structures (i.e., $\gamma$=1/2 describes commonly low temperature dc conductivity in carbon nanotubes) [47]. The spherical shape of CNOs favors 3D hopping occurs, in accordance with experimental observations in fullerenes [48], for which a $T^{-1/4}$ dependence (3D hopping) was determined [49],

$$\sigma_{0,VRH}(T) = A_{VRH} \exp(-(T_{0,VRH}/T)^{1/4}) \qquad (15)$$

$T_{0,VRH}$ is a constant proportional to the density of states at the Fermi level $N(E_F)$ and $T_{VRH0} = C/(L^3 k N(E_F))$ where L denotes the radius of the localized wave function decay of the charge carriers [49 - 52], q is the electron's electric charge and C is a dimensional constant ($C \approx 18.1$). $A_{VRH}$ is assumed to be practically constant. The average hopping length is $R = [9L/(8\pi k T N(E_F))]^{1/4}$ . . The last two relations, combine to:

$$R = L[9T_{0,VRH}/(8C\pi T)]^{1/4} \qquad (16)$$

Eqs. (7) and (15) can be re-written as:

$$f_{0,\sigma,VRH}(T) = (A_{VRH}/(2\pi\varepsilon_\infty))\exp(-(T_{0,VRH}/T)^{1/4}) \qquad (17)$$

where $A_{0,\sigma,VRH} \equiv A_{VRH}/(2\pi\varepsilon_\infty)$. Note that while GM and VRH involve quantum mechanical tunneling assisted by phonons, the first extend electron states existing in mesoscopic 'metallic' grains, while the latter mechanism refers to localized states centered at the different positions in the band gap.

$Ln\sigma_0$ and $lnf_{0,\sigma}$ are plotted against 1/T in Figures 10 and 11, respectively. The fitting functions are:

$$\sigma_0(T) = \sigma^* + \sigma_{0,VRH}(T) + \sigma_{0,GM}(T) \quad (18)$$

where $\sigma^*$ is assumed to be temperature independent term (related to a coherent tunneling), and:

$$f_{0,\sigma}(T) = f_{0,\sigma}^* + f_{0,\sigma,VRH}(T) + f_{0,\sigma,GM}(T) \quad (19)$$

where $f_{0,\sigma}^*$ denotes a temperature independent term. The fitting procedure indicated that the constant terms $\sigma^*$ and $f_{0,\sigma}^*$ are zero within fitting errors. The fitting parameters of the dc conductivity equation (Eq. (18)) are: $A_{GM} = (5.3 \pm 0.6) \times 10^{-4}$ S/cm, $T_{0,GM} = (3300 \pm 400)$K, $A_{VRH} = (10 \pm 4) \times 10^{-6}$ S/cm and $T_{0,VRH} = (220 \pm 80)$K. For the conductivity relaxation equation (Eq. (19)) the values are: $A_{0,\sigma} = (2.1 \pm 0.6) \times 10^6$ Hz, $T_{0,GM} = (3000 \pm 400)$K, $A_{0,\sigma,VRH} = (9 \pm 4) \times 10^4$ Hz and $T_{0,VRH} = (270 \pm 80)$K. Note that apart from the $T^{-1/4}$ dependence for the VRH term, used in the above mentioned fitting, we also checked for $T^{-1/2}$ or $T^{-1/3}$ dependencies, but the fitting was rather poor, compared with that when 3D hopping was assumed.

According to the granular metal (GM) model, the ratio of the average separation spacing s over the mean diameter of a conducting grain can be estimated through eq. (13). By replacing in eq. (13) the values of $T_{0,GM}$ obtained from the fitting proceduer, U=2 eV we get a n $s/d \cong 0.1$. The diameter d of CNOs ranges from 8 to 20 nm [7[, thus, the effective grain separation d ranges from about 0.8 to 2 nm. Furthermore, the fitting parameters of the VRH model can yield an estimate of the mean hopping distance of electric charge carriers, according to eq. (16). Using the fitting parameter $T_{0,VRH}$ found above, setting C=18.1 and taking the length constant of electron function to be of the order of the CNO diameter L ranging from 8 to 20 nm [7], we get hopping distance values ranging from 4 -10 nm at room temperature.

Note that, when a fractional exponent is smaller than 1, relaxation time shows deviations from Arrhenius as a Vogel-Fulcher-Tamman (VFT) for involving atomic or molecular scale relaxation [52 - 54]. Since electron conducting polymers exhibit an opposite curvature than that predicted from VFT, its modification based on physical arguments would stimulate research on the probable extension of a novel generalized VFT law, that can describe relaxation in organic semi-conductors.

## 4. Conclusions

The temperature dependence of the complex electric modulus and the ac conductivity of CNO/PANI composites, studied from 1 mHz to 1 MHz under isothermal conditions ranging from 15 K to room temperature, indicate two phonon assisted quantum mechanical tunneling processes. One is associated with inter-grain hopping of extended electronic states in mesoscopic highly conducting PANI grains hosted in a less conducting PANI environment, while, the other is ascribed to 3D hopping within CNOs and clusters. Electric modulus and dc conductivity analyses indicate a crossover temperature at 115 K. The ac conductivity fractional exponent studies and the bifurcating scaling of these isotherms suggested that this transition is roughly around 115 K, respectively. Based on the present results, the electrical properties of the CNOs/PANI can be optimized further for usage as super-capacitor electrodes by tuning the individual conductivity mechanisms with proper doping and structural modification of the composites.

**cknowledgements**

We gratefully acknowledge the financial support of the National Science Centre, Poland (grant No.2012/05/E/ST5/03800) to M.E.P.-B. L.E. thanks the Robert A.

Welch Foundation for an endowed chair, grant #AH-0033 and the US NSF, grants: DMR-1205302 and CHE-1408865.

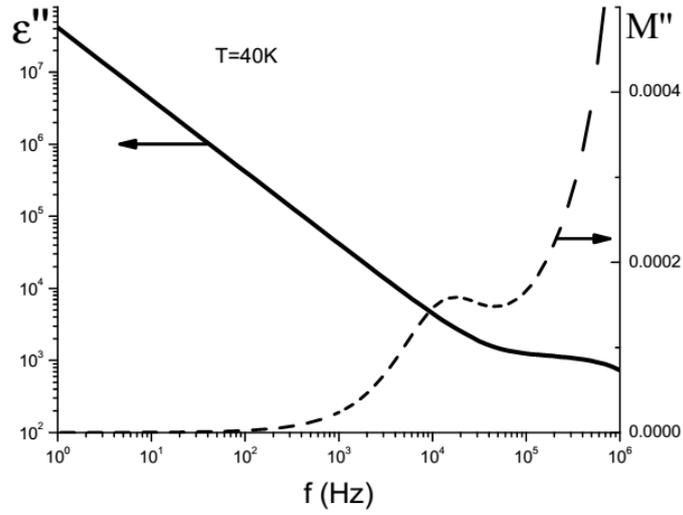

Figure 1: Isotherms of ε´´ (solid line) and M´´ (dash line) vs. frequency recorded at 40 K.

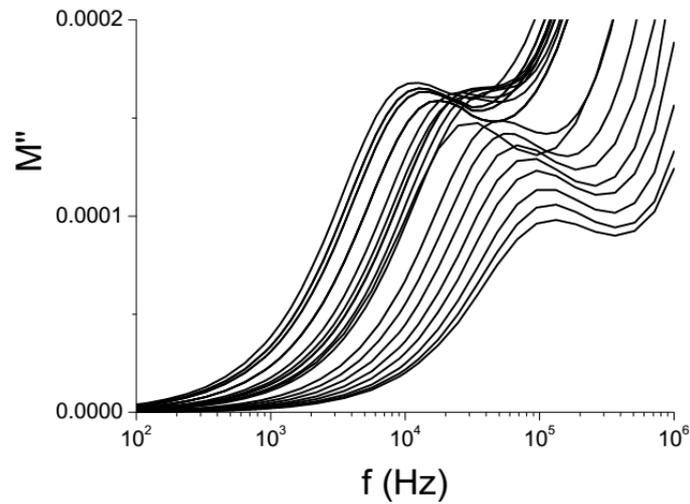

Figure 2: Isotherms of M´´ vs. f, recorded (from left to right) at 15, 23, 30, 40, 50, 60, 70, 80, 90, 100, 110, 130, 150, 170, 190, 210, 230, 250, 270 and 293 K, respectively.

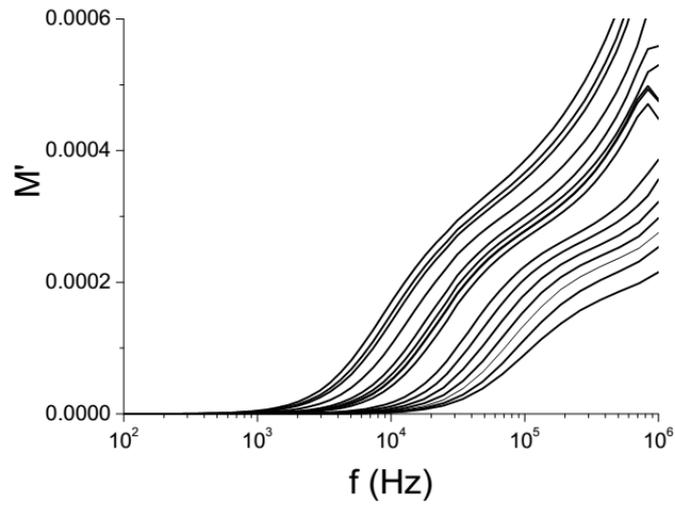

Figure 3: Isotherms of M´ vs. f, recorded (from top to bottom) at 15, 23, 30, 40, 50, 60, 70, 80, 90, 100, 110, 130, 150, 170, 190, 210, 230, 250, 270 and 293 K, respectively.

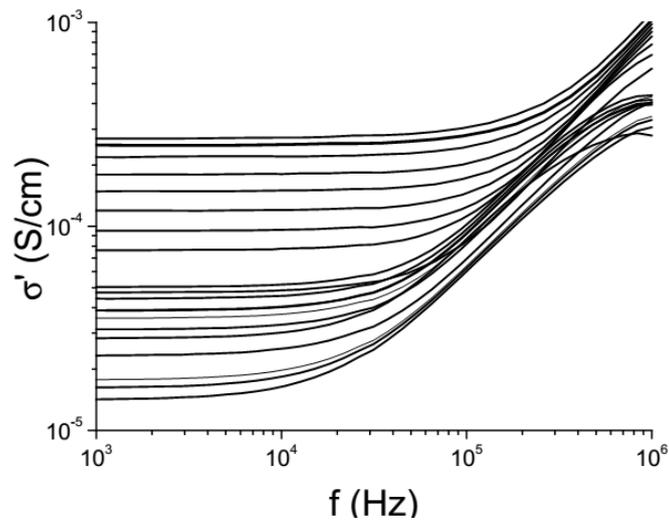

Figure 4: The real part of the ac conductivity σ´(ω), recorded isothermally at the same temperatures depicted in Figure 3. (bottom line: 15 K; upper line: room temperature).

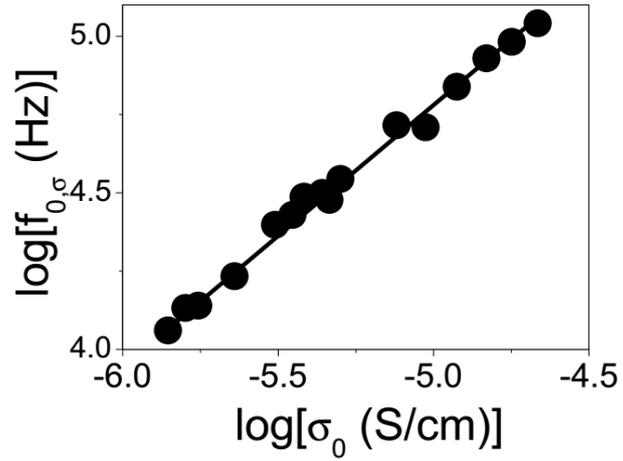

Figure 5: The logarithm of the conductivity relaxation frequency obtained from the maxima of M´´(f) isotherms vs. the logarithm of the dc conductivity determined by fitting eq. (9) to σ´(f) isotherms. The straight line best fits the data points.

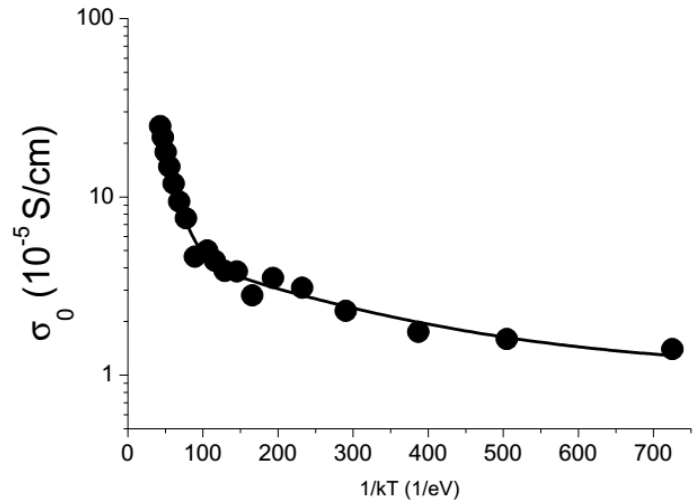

Figure 6: $\sigma_0$ vs $1/kT$. The line is the best fit of Eq. (11) to the data points.

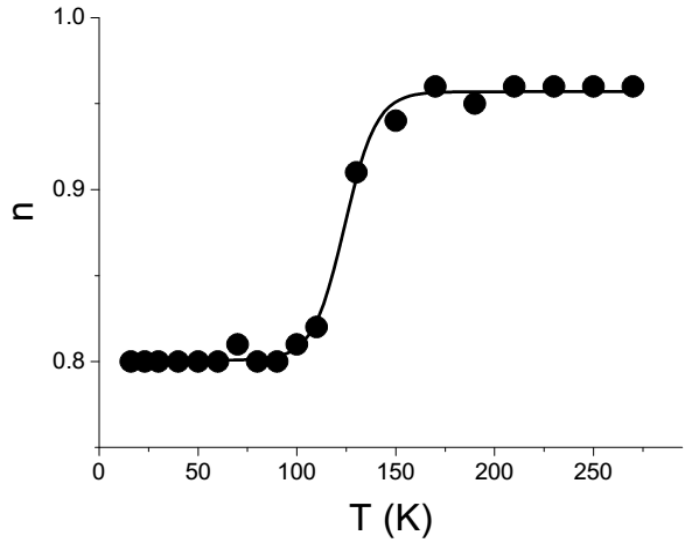

Figure 7: The fractional exponent n of the power-law dispersive ac conductivity described by Eq. (7) vs. the temperature where isotherms of σ´(f) were recorded.

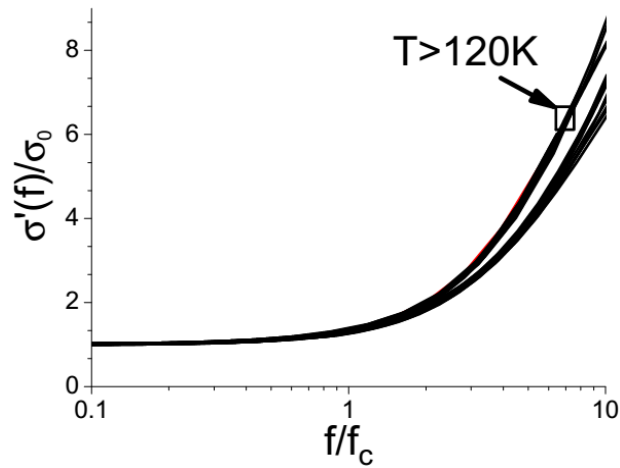

Figure 8: Scaling of σ´ isotherms depicted in Figure 4. The critical frequency $f_c$ is determined according to Almond and West. Isotherms above and below 115 K superimpose on two distinct master curves, respectively, which 'bifurcate' at high frequencies, indicating different electric charge transport dynamics.

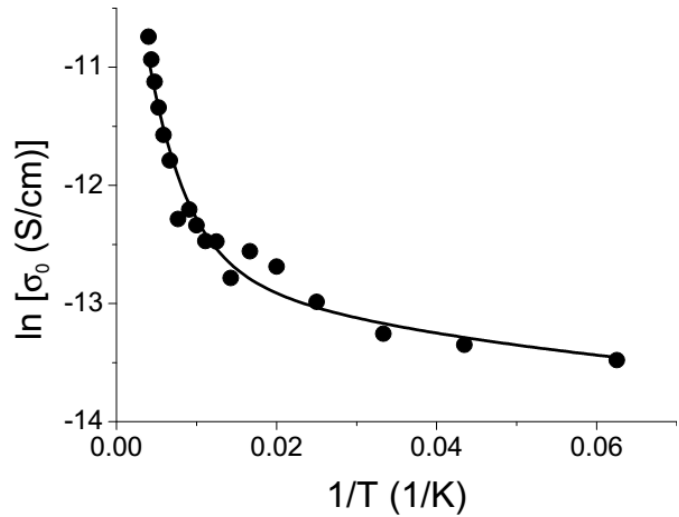

Figure 9: The natural logarithm of the dc conductivity σ0, determined from the fitting Eq. (7) to the σ´(f) data, as a function of reciprocal temperature. Solid line is the best fit of Eq. (17) to the data points.

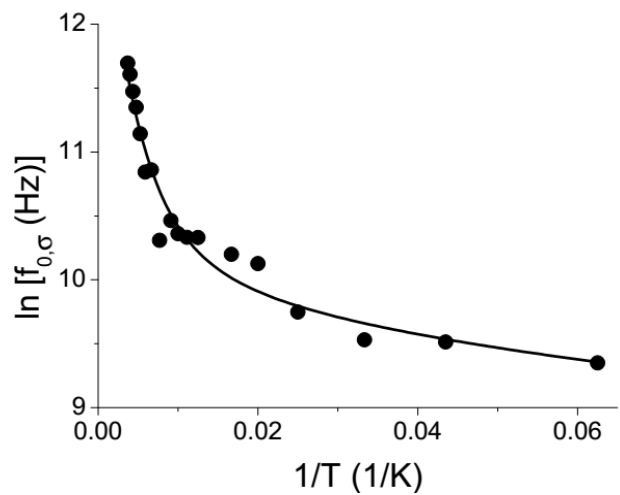

Figure 10: The natural logarithm of the relaxation conductivity frequency $f_{0,\sigma}$ vs. reciprocal temperature. Solid line is the best fit of Eq. (19) to the data points.